\documentclass[11pt]{article}

\usepackage{latexsym,amssymb,amsmath,amscd,amsthm,amsxtra}
\usepackage{amsmath}
\usepackage{amsfonts}
\usepackage{amssymb}
\usepackage{graphicx}%
\usepackage[margin=1.5in]{geometry}
\usepackage{graphicx}
\usepackage{amssymb, amsmath}
\usepackage{epstopdf}
\title{Frobenius transformation, mirror map and instanton numbers}
\author{Albert Schwarz, Vadim Vologodsky}
\begin{document}
\maketitle

{\bf Abstract}

We show that one can express Frobenius
transformation  on middle-dimensional $p$-adic cohomology of Calabi-Yau threefold in terms of mirror map
and instanton numbers.
We  express the mirror map 
in terms of Frobenius transformation on
$p$-adic cohomology .  We discuss  a $p$-adic  interpretation
of the conjecture about integrality of Gopakumar-Vafa invariants.

\section{Introduction}
In [1] we gave an expression of instanton numbers in terms of Frobenius  transformation on middle-dimensional $p$-adic
cohomology  of  Calabi-Yau threefold
with respect to canonical coordinates
on moduli space of complex structures  . We mentioned that conversely Frobenius  map  in canonical coordinates can be expressed in terms of instanton numbers. (A rigorous exposition
of the results of [1] was given in [2].)
In present paper we will prove that Frobenius map on middle-dimensional cohomology constructed by means of any coordinate system in the neighborhood of maximally unipotent boundary point of complex structures can be expressed in terms of mirror map and instanton numbers (Sec.2) The calculations in Sec. 2  generalize some calculations of [1],
however, our  exposition is  independent of [1].
In Sec. 3 we will  discuss integrality of mirror map ; we will give an explicit formula for mirror map
in terms of Frobenius transformation  on cohomology
of Calabi-Yau manifold. (Talking about Frobenius map on cohomology we have in mind the middle-dimensional cohomology with coefficients in the ring $\mathbb{Z}_p$ of
integer $p$-adic numbers; see  [3] for transparent construction of this map.) In Sec.4 we interpret the conjecture about  integrality of Gopakumar-Vafa invariants in terms of Frobenius map on $p$-adic wave functions.
We hope that this interpretation will allow us to prove the conjecture.  In the last section we describe a way that permits us to apply our calculation of Frobenius map to  computation of zeta-functions of Calabi-Yau manifolds over finite fields. These zeta-functions were studied in [8]; our calculations are more general , but  less explicit.

\section{Calculation of Frobenius map}
Let us consider a family of complex 
Calabi-Yau manifolds of dimension $n$ in the neighborhood of maximally unipotent boundary point in the moduli space $\mathcal{M}$ of complex structures. The coordinates in the neighborhood of the boundary point will be denoted by $z_1,...,z_r$; the points on the boundary divisor  obey $z_i=0$. 
We assume that the moduli space  $\mathcal{M}$ is defined over $\mathbb{Z}.$

The corresponding B-model can be formulated in terms of Hodge filtration $\mathcal{F}^p$ , monodromy
 weight filtration  $\mathcal{W}_k$ and Gauss-Manin connection $\nabla$ on the bundle of cohomology groups $H^n$ (with complex coefficients). (See, for example, Sec 5.1 of [4] .) We assume that Morrrison integrality conjecture (see [5] or Sec 5.2.2 of [4] ) is satisfied.

We will consider Calabi -Yau threefolds. In  this case the  cohomology group $H^3$ can be represented as a direct sum
of the subgroups $I^{k,k}=\mathcal{W}_{2k}\bigcap\mathcal{F}^k$ where
$k=0,1,2,3,$ and rank$I^{0,0}=$rank$I^{3,3}=1,$rank$I^{1,1}=$
 rank$I^{2,2}=r.$
 ( Here $r$ stands for the dimension of
$\mathcal{M}$.)

Let us assume for simplicity that $r=1.$ Then we can take a symplectic basis of $H^3$ consisting of  vectors $e_k\in I^{3-k,3-k},\ ,k=0,1,2,3.$  ( In other words, the vectors $e_k$ obey $<e_3,e_0>=1, <e_2,e_1>=-1.$)

For appropriate choice of this basis (see, for example [4]) Gauss- Manin connection takes the form
\begin{equation}
\label{gm1}
\nabla _{\delta} e_3=0,
\end{equation}
\begin{equation}
\label{gm2}
\nabla _{\delta} e_2=Y_3(z)e_3,
\end{equation}
\begin{equation}
\label{gm3}
\nabla _{\delta} e_1=Y_2(z)e_2,
\end{equation}
\begin{equation}
\label{gm4}
\nabla _{\delta} e_0=Y_1(z)e_1.
\end{equation}
Here $\delta$ stands for logarithmic derivative with respect to
coordinate $z$ in the neighborhood of maximally unipotent boundary point. Compatibility
of Gauss-Manin connection with symplectic structure on cohomology  implies that $Y_1=Y_3.$

We follow the notations of Sec 5.6.3 of [4]. It is shown in this section that the mirror map $q(z)$ can be expressed in terms of $Y_1=Y_3$; namely $\delta \log q= Y_3.$ In canonical coordinate  $q$ we have $Y_1=Y_3=1$ (for appropriate normalization of the basis in cohomology) and $Y_2$ can be expressed in terms of instanton numbers (up to a constant summand.) In general  $Y_i$ can be expressed in terms of mirror map and instanton numbers. There exists also an expression of $Y_i$ in terms of  solutions of Picard-Fuchs equations (of periods of holomorphic
3-form $\Omega =f e_0.$) We will need the relation
\begin{equation}
\label{fyy}
\delta \log (f^4Y_1^3Y_2^2Y_3)=c_3,
\end{equation}
 where $c_3$ stands for one of the coefficients of Picard-Fuchs equation
 \begin{equation}
\label{ }
\nabla _{\delta} ^4 \Omega=c_3 \nabla _{\delta} ^3\Omega+c_2 \nabla _{\delta} ^2 \Omega+c_1 \nabla _{\delta} \Omega +c_0 \Omega.
\end{equation}
To obtain (\ref {fyy}) we substitute $\Omega =f e_0$ into Picard-Fuchs equation and apply (\ref{gm1})-(\ref{gm4}).  We will use (\ref {fyy}) to calculate $f^2Y_1^2Y_2.$

One can prove that the action of Gauss-Manin connection
on cohomology with coefficients in $\mathbb{Z}_p$ is given by
(\ref{gm1})-(\ref{gm4}) where $Y_i$ are considered as series with
$p$-adic coefficients ;see [2]. (In this statement  one should  assume that
the canonical coordinate and the vectors $e_i$  are normalized in appropriate way.)

The map $z\to z^p$ is called Frobenius map;
it induces a map on functions of $z$ denoted by
${\rm Fr}$. It is a non-trivial fact that it can be lifted to a map of cohomology groups 
with coefficients in $\mathbb{Z}_p$; this map  (Frobenius transformation) also will be denoted by ${\rm Fr}$. Frobenius transformation is compatible with symplectic structure on cohomology:
\begin{equation}
\label{cf}
<{\rm Fr} a, {\rm Fr} b >= p^3 {\rm Fr}<a, b>,
\end{equation}
 where $<a, b>$ stands for the inner product
of cohomology classes.

The Frobenius transformation does not preserve the Hodge filtration; however, for $p>3$ it satisfies
\begin{equation}
\label{ch}
{\rm Fr}\mathcal{F}^s\subset p^s\mathcal{F}^0.
\end{equation}
{\it We will always impose the condition $p>3.$}

The Frobenius map preserves the monodromy weight filtration;
hence its matrix is triangular in our basis:
\begin{equation}
\label{Fr1}
{\rm Fr}e_3=a_{33}e_3,
\end{equation}
\begin{equation}
\label{Fr2}
{\rm Fr}e_2=pa_{22}e_2+pa_{32}e_3, 
\end{equation}
\begin{equation}
\label{Fr3}
{\rm Fr}e_1=p^2a_{11}e_1+p^2a_{21}e_2+p^2a_{31}e_3,
\end{equation}
\begin{equation}
\label{Fr4}
{\rm Fr}e_0=p^3a_{00}e_0+p^3a_{10}e_1+p^3a_{20}e_2+p^3a_{30}e_3.
\end{equation}

Notice that in the LHS of these formulas the vectors $e_i$ are calculated
at the point $z$ and in the RHS these vectors are calculated at the point $z^p.$
The coefficients belong to $\mathbb{Z}_p[ [z]].$

From the relation (\ref{cf}) we obtain
\begin{eqnarray}
      &a_{33}a_{00}=1,a_{22}a_{11}=1, a_{00}a_{32}=a_{22}a_{10},    \\
      &  a_{11}a_{20}-a_{21}a_{10}+a_{31}a_{00}=0 
\end{eqnarray}
The Frobenius map is compatible with Gauss-Manin connection:
\begin{equation}
\label{c}
\nabla _{\delta}{\rm Fr}=p{\rm Fr}\nabla _{\delta}.
\end{equation}
Using (\ref{c}) we obtain relations between $Y_i$ and  $a_{ij}:$
\begin{equation}
\label{r}
\delta a_{ji}+a_{j-1,i}Y_j=a_{j,i+1}{\rm Fr}(Y_{i+1}).
\end{equation}
We conclude that
\begin{equation}
\delta a_{ii}= 0 
\end{equation}
\begin{equation}
\delta a_{i,i-1}=a_{i,i}{\rm Fr}Y_i- a_{i-1,i-1}Y_i
\end{equation}
\begin{equation}
\delta a_{i,i-2}=a_{i,i-1}{\rm Fr} Y_{i-1}-a_{i-1,i-2}Y_i
\end{equation}
\begin{equation}
\delta a_{30}=a_{31}{\rm Fr}Y_1-a_{20}Y_3.
\end{equation}

It follows from these formulas that $a_{i,i}$ does not depend on $z$ and $i$. ( Independence of $z$ follows from the first equation;
independence of $i$ follows from second equation and from the remark that LHS of  all equations vanish at $z=0.)$ Let us denote the common value of $a_{ii}$ by $\epsilon;$ we obtain from (13) that
$\epsilon=\pm 1.$  

The above formulas immediately permit us to calculate every matrix element of Frobenius map up to a constant summand.  To fix these constants it is sufficient to know the matrix of Frobenius map  at the point  $z=0.$ From  (\ref {r})
we can  deduce that vanishing one of the elements of Frobenius 
matrix at $z=0$ implies that all other elements on the same diagonal (with the same difference between indices) are zero.
From the other side it is proved in [2] that in canonical coordinates $q$ we have $a_{12}=0$ ; this means that in any integral coordinate system $z$  
we have   $a_{12}=0$ at $z=0 .$ (This follows from remark that
$dq/dz=\pm 1$ at $z=0.)$  Applying (12),(13) we conclude that
all off-diagonal entries of Frobenius matrix at $z=0$ vanish,
except, may be, $a_{30}.$ 

Similar calculations can be used to calculate the Frobenius map for $r>1.$ In this case with an appropriate choice of symplectic basis   $e_0, e_1,...,e_r, e^0,e^1,...,e^r$ where $e_0\in I^{3,3},
e_1,...,e_r\in I^{2,2}, e^1,...,e^r\in I^{1,1}, e^0\in I^{0,0},$
Gauss-Manin connection  can be represented in the following way
\begin{equation}
\label{GM}
\nabla _{{\delta _i}}e^0=0,\\
\nabla _{{\delta _i}}e^j=^3Y_i^je^0,\\
\nabla _{{\delta _i}}e_j=^2Y_{ijk}e^k,\\
\nabla _{{\delta _i}}e_0=^1Y_i^je_j.
\end{equation}
We use the notation $\nabla _{{\delta _i}}$ for the covariant derivative corresponding to $\delta _i=z_i\partial/\partial z_i.$
In canonical coordinates $^1Y_i^j=^3Y_i^j=\delta _i^j,$ hence in arbitrary coordinate system one can express $^1Y_i^j=^3Y_i^j$ 
in terms of mirror map. (See [4],
Sec. 5.6.3 or [7].)  Taking into account that in canonical coordinates $^2Y_i^j$ is the normalized Yukawa coupling that
can expressed in terms of instanton numbers we see that 
all coefficients in (\ref {GM}) can be expressed in terms of mirror map and instanton numbers. From the other side repeating the considerations used for $r=1$ we relate these coefficients to the Frobenius map.

Notice that we can consider also a transformation of cohomology groups 
over $z$ into cohomology groups over $z^q$ where $q=p^s$. This map
denoted by ${\rm Fr}^{(s)}$  is induced by the $s$-th power of map $z\to z^p.$
One can calculate this map as a composition of $s$ Frobenius transformations.
From the other side we can use the compatibility of this map with Gauss- Manin
connection to calculate it in the same way as  Frobenius map.
Composing Frobenius  transformations we obtain
\begin{equation}
\label{frs}
{\rm Fr}^{(s)}e_i=\sum p^{(3-i)+(s-1)(3-j)}a_{ji}^{(s)}e_j
\end{equation}
where  $j\geqslant i$ and $a_{ii}=\epsilon ^s.$
The equation for the coefficients in (\ref {frs}) has the form
\begin{equation}
\label{frsc}
p^{s-1}a_{i+1,j}^{(s)}{\rm Fr}^{(s)}Y_{i+1}=\delta a_{ij}^{(s)}+a_{i,j-1}^{(s)}Y_j
\end{equation}

\section{Expression for mirror map}

Integrality of mirror map was proven in  [6] for quintic and in [2] in
general case. In this section we will use the ideas of [2] to give an expression for mirror map in terms of Frobenius transformation.
Integrality of mirror map will follow from this expression. 

Our main tool  will be the Artin-Hasse exponential
function
$$E_p(x)=\exp (x+\frac{x^p}{p}+\frac{x^{p^2}}{p^2}+...).$$
The expansion of this function with respect to $x$ has integer
$p$-adic coefficients: $E_p(x)\in \mathbb{Z} _p [[x]].$
It is easy to check that
$$\delta \log E_p(x^{pk}) -\delta \log E_p(x^k) =-kx^k=-\delta x^k.$$
Using this identity we can derive the following 

{\bf Lemma  1}

For every  $r(x)\in \mathbb{Z}_p[[x]]$ one can find
such a series $Q(x)\in x\mathbb{Z} _p [[x]]$ that
$\nu (x) =\delta \log Q(x)$ obeys
$$\nu (x^p)-\nu (x)=\delta r(x). $$

To prove this lemma we give an explicit construction of $Q(x)$ in 
terms of Artin-Hasse exponential. Let us suppose that 
$r(x)=\Sigma r_kx^k .$ Then it follows immediately from the
properties of Artin-Hasse exponential that  the power expansion of the product
$Q(x)=x\Pi_kE_p(x^k)^{-r_k}$  is a series we need.

 Artin-Hasse exponential can be expressed as a product:
 $$E_p(x)=\Pi (1-x^n)^{-\frac{\mu (n)}{n}}$$
where $n$ runs over all natural numbers that are not divisible by $p$ and $\mu$ stands for Moebius function. Using this expression we can obtain another representation of $Q(x):$
\begin{equation}
\label{ }
Q(x)=x\Pi_k (1-x^k)^{s_k}
\end{equation}
where $$s_k=\Sigma\frac{\mu (d)}{d}r_{\frac{k}{d}}$$ and $d$ runs
over all divisors of $k$ that are not divisible by $p.$

In the situation considered in Sec. 2 integrality of mirror map 
and an explicit expression for it immediately follows from this lemma applied to the function $r(z)=\pm a_{32}(z).$ 
Using (18) and taking into account that $a_{ii}=\epsilon=\pm 1$ 
we obtain $$\delta a_{i,i-1}=\pm ({\rm Fr}Y_i- Y_i).$$
Applying this relation for $i=3$
we can say that the role of the function $\nu$ is played by the function $Y_3.$ The mirror map $q$ obeys $\delta \log q=Y_3$;
this equation specifies $q$ up to a constant factor. We come to a conclusion that
\begin{equation}
\label{ }
q(z)=const z\Pi_kE_p(z^k)^{\mp a_{32}^k}
\end{equation}
where $a_{32}^k$ denotes the coefficient of $z$-series $a_{32}(z).$

One can prove that in appropriate normalization of mirror map
the constant in this formula is equal to $\pm 1.$ We will not discuss the calculation of this constant referring to [2].

Applying  (23) we can present the expression for $q(z)$ in the form
$$q(z)=const z(1-z^k)^{\sigma _k},$$
where
$$\sigma _k=\sum \frac{\mu (d)}{d} \epsilon a_{32}^\frac{k}{d}.$$

Notice that applying the above considerations to the relation (18) for arbitrary $i$
we can prove integrality
of all functions $q_i (z)$ obeying  
\begin{equation}
\label{q}
\delta \log q_i=Y_i.
\end{equation}
 These functions generalizing the mirror map were introduced by Deligne  [7]  in geometric way; see also [4],Sec. 5.6.3. Notice that 
  the equation (\ref{q}) specifies  the function $q_i (z)$  only up to a constant factor, but the geometric definition fixes this factor.

It follows from the proof of integrality of $q_i (z)$  that
\begin{equation}
\label{ }
a_{i, i-1}=\pm \log \frac{q_i (z^p)^{\frac{1}{p}}}{q_i (z)} + const
\end{equation}

One can prove that the constant in this equation vanishes.

We have considered the mirror map  only for Calabi-Yau threefolds  and only in the case when $r=1.$ However, our consideration can be applied also to $n$-dimensional Calabi-Yau manifolds without any restriction on the dimension $r$ of moduli space of complex structures.

In the case $r>1$  one should apply the following multidimensional generalization of Lemma 1.

{\bf Lemma 2}

Let us suppose that  $r^i(z)\in \mathbb{Z}_p[[z_1,...,z_r]].$ Then one can find
such  series $Q^i(z)\in z_i\mathbb{Z} _p [[z_1,...,z_r]]$ that
$\nu _j^i =\delta _j \log Q^i$ obeys
$${\rm Fr}\nu_j^i -\nu _j^i=\delta _j r^i.$$
Here $\delta_j$ stands for the logarithmic derivative with repect to $z_j$ and the Frobenius map ${\rm Fr}$ transforms $f(z_1,...,z_r)$ into $f(z_1^p,...,z_r^p).$

Again we can write  
$$Q^i(z)=z_i\Pi_kE_p(z^k)^{-r_k^i},$$
where $r_k^i$ are coefficients of the power expansion of $r^i(z)$
(we consider $k$ as multiindex).
\section{Free energy and wave function}
Let us consider  B-model in canonical coordinate $q$ on the moduli space $\mathcal{M}$  of complex structures. (We assume
that the dimension $r$ of $\mathcal{M}$ is equal to 1, but the generalization to the case $r >1$ is straightforward.)
Then the calculations of Sec. 2 coincide with the calculations of [1] and we obtain the following expression for Frobenius map:

\begin{equation}
\label{Frr1}
{\rm Fr}e_3=e_3,
\end{equation}
\begin{equation}
\label{Frr2}
{\rm Fr}e_2=pe_2, 
\end{equation}
\begin{equation}
\label{Frr3}
{\rm Fr}e_1=p^2e_1+p^2a_{21}e_2+p^2a_{31}e_3,
\end{equation}
\begin{equation}
\label{Frr4}
{\rm Fr}e_0=p^3e_0+p^3a_{20}e_2+p^3a_{30}e_3,
\end{equation}
where all coefficients can be expressed in terms of
$a=a_{30}$; namely $$a_{21}=-\frac{1}{2}\delta ^2 a,a_{31}=- a_{20}=\delta a.$$
Here we consider the case when all diagonal coefficients are equal to $1;$ in the case when the diagonal coefficients are equal to $-1$ we should change  some signs. The symbol $\delta$ stands for logarithmic derivative with respect to $q$ (for derivative with
respect to $t=\log q$).

In canonical coordinates  the function $Y_1=Y_3$ is equal to $1$
and the function $Y=Y_2$ is the Yukawa coupling (third logarithmic derivative of genus zero free energy $f=F_0).$
If the Yukawa coupling has the form
$$Y(q)=c+\sum n_kq^k,$$
the  genus zero free energy  can be represented as
$$f(q)=ct^3+\sum k^{-3}n_kq^k.$$
It was derived in [1] from ( \ref {Frr1}-\ref {Frr4}) and compatibility 
of Frobenius map with Gauss- Manin connection (15) that
\begin{equation}
\label{Y}
Y(q^p)-Y(q)=\frac{1}{2}\delta ^3 a.
\end{equation}
(The equations (\ref {Frr1}-\ref {Frr4})  and (\ref {Y})  follow from (17)-(20).)
For genus zero free energy $f=F_0$ we obtain
\begin{equation}
\label{f}
p^{-3}f(q^p)-f(q)=\frac{1}{2} a.
\end{equation}

It was shown in [1] that (\ref {Y}) can be used to analyze integrality
of instanton numbers.
It was mentioned there that the main lemma that permitted us
to give $p$-adic interpretation of instanton numbers 
(=genus zero Gopakumar-Vafa invarians) can be generalized
to the case of Gopakumar-Vafa invariants for arbitrary genus.
Let us formulate now one of possible statements of this kind.

Recall  first of all that  the free energy
\begin{equation}
\label{F}
F(q)=\sum \lambda ^{2g-2} F_g(q)
\end{equation}
can be represented as a sum of the contribution of non-trivial
instantons $F'$ and the contribution of constant maps $F''.$ The first summand can be expressed in terms of Gopakumar-Vafa invariants
$n^g_k$:
$$F'=\sum n^g_k\frac{1}{m}(2\sin \frac{m\lambda}{2})^{2g-2}q^{mk}.$$
We sum here over all possible values of  $g=0,1,2,...$ and over 
$k,m\in \mathbb{N}.$
The second summand has the form
$$F''=\lambda ^{-2}\frac{ct^3}{6}-\frac{dt}{24}$$
where $c$ and $d$ are integers. Using these formulas one can represent the partition function
$Z$ as $Z' Z''$ where
\begin{equation}
\label{Z}
Z'=\exp (F')=\prod (1-\Lambda ^rq^k)^{m^r_k}
\end{equation}
\begin{equation}
\label{z}
Z''=\exp (F'').
\end{equation}
Here $\Lambda = e^{-i\lambda},$ $$m^r_k=rn^0_k+\sum\limits_{g\geq {1+|r|}} (-1)^{g-1}n^g_k \left(\begin{array}{l}2g-2\\
g-1-r\end{array}\right ).$$
 (In the formulas for free energy we disregard a constant summand, in the formulas for partition function we disregard a constant factor.)

The derivation of these formulas is based on elementary identities
$$\exp {\big (\sum _m\frac{1}{m}\big(2\sin \frac{m\lambda}{2}\big)^{2g-2}q^{mk}\big )}=\prod _{r=0}^{2g-2}\big (1-\Lambda ^{k-g+1}q^n\big)^{
{(-1)}^{r+g}\binom {2g-2}{r}}$$
for $g>0,$
$$\exp {\big (\sum _m\frac{1}{m}\big(2\sin \frac{m\lambda}{2}\big)^{-2}q^{mk}\big )}=
\prod _{r=1}^\infty \big (1-\Lambda ^{-r}q^n\big)^r$$
for $g=0.$ ( See  [13].) 

The 
integrality of numbers $m^r_k$ follows from integrality of Gopakumar-Vafa invariants. Conversely, one can prove that the integrality of the numbers $m^r_k$ together with integrality of 
instanton numbers $n^0_k$ implies the integrality of all Gopakumar-Vafa invariants $n^g_k.$ Further, the integrality of  $m^r_k$ is equivalent  to the integrality of $Z'$ considered as an element of the ring of power series with respect to $q$ with coefficients in Laurent series with respect to $\Lambda.$ (The integrality is understood as integrality of coefficients of Laurent series.)  The integrality
of $Z'$ can be expressed also in terms of $p$-adic reduction of
$Z'.$ ( We can consider $p$-adic reduction of it because the coefficients in series with respect to $q,\Lambda$ and, in the case, of $Z''$ with respect to $t=\log q, i\lambda =\log \Lambda $ are rational.) It is easy to check that for this reduction
\begin{equation}
\label{i}
\frac{Z'(\Lambda ^p, q^p)}{Z'(\Lambda, q)^p}=1+p\zeta(\Lambda, q), 
\end{equation}
where $\zeta$ is a power series
with respect to $q$ having as coefficients Laurent series with 
respect to $\Lambda$ ; the coefficients of these Laurent series are $p$-adic integers. Notice that 
\begin{equation}
\label{F''}
F''(p\lambda, pt)-pF''(\lambda, t)=0,
\end{equation}
hence
$$\frac{Z''(p\lambda, pt)}{Z''(\lambda,t)^p}=1.$$
We see that $Z$ also has  integrality properties
generalizing (\ref {i}).

One should emphasize that the formula (\ref {Z}) is valid only  for
$|\Lambda|>1$ (we used this condition in the  analysis of  the contribution of $g=0.$)  However, we can use the fact that $Z$ is
a function   that is even with respect to $\lambda$
(i.e. it  does not change under the substitution $\Lambda \to \Lambda ^{-1}$); therefore it can be considered as a function of
$\nu= \Lambda +\Lambda ^{-1}-2 =-2\sin ^2\frac{\lambda}{2}$
having a first order pole at $\nu =0.$ It is easy to derive from the above statements that the Laurent series  with respect to $\nu$
that appear in $Z'$  as coefficients in the power series with respect to $q$ have integral coefficients. The derivation is based
on the existence of one-to-one correspondence   between
integer polynomials with respect to $\nu$ and integer 
Laurent polynomials with respect to $\Lambda$ that do not change under the substitution $\Lambda \to \Lambda ^{-1}.$
It uses also the fact that Laurent series for $\nu ^{-1}$ at the point $\Lambda =\infty$ contains only negative powers of $\Lambda$ appearing with integer coefficients.

Notice that the integrality properties can be expressed  in terms of $p$-adic reduction of free energy. 
Namely, it follows from the above results that 
\begin{equation}
\label{ }
F'(p\lambda, q^p)-pF'(\lambda, q)=p \tau ,
\end{equation}
where $\tau \in \nu ^{-1}\mathbb{Z}_p[[\nu,q]].$ 
This statement was mentioned in [1],[2].  
Using (\ref {F''}) we obtain that $F=F'+F''$  has the same integrality property  as $F'.$

B-model for arbitrary genus  ( or, more precisely, corresponding
topological string)  can be obtained from genus zero B-model by means of quantization [11]. 
The wave function of B-model in holomorphic polarization (in polarization corresponding to the basis $e_k$) can be expressed in terms of
free energy (\ref {F}) in the following way
\begin{eqnarray}
\Psi&=\exp\big(F(\frac{\lambda}{\rho}, t+\frac{x}{\rho})-\lambda^{-2}(F_0(t)\rho^2+\frac{\partial
F_0}{\partial t}x\rho+\frac
{1}{2}\frac{\partial^2
F_0}{\partial t^2}x^2)\nonumber\\
&-(\frac{\chi}{24}-1)\log(\rho)\big).
\end{eqnarray}
where  $(x,\rho)$ are coordinates on Lagrangian subspace spanned by $e_1,e_0$ and $\chi$ is an integer.
(This expression was derived in [12] from the results of
[10].)  Let us restrict our attention to the class of wave functions
having the form $\psi=e^f$ where $f=\hbar ^{-1} f_0 +f_1+\hbar f_2+...$ (to the class of semiclassical wave functions). In B-model $\hbar=\lambda ^2$; the wave function of B-model belongs to our class. It is easy to check that a semiclassical wave function remains in the same class when we change  the polarization
(i.e.  the group of symplectic transformation acts projectively on our space). (See [12], the end of Sec 2.) Moreover, the formulas
relating the wave function $\psi =e^f$ to the wave function 
$\psi '=
e^{f'}$ give a rational expression of $f'$ in terms of $f$, hence they can be applied over any field, in particular over a field of $p$-adic numbers. This remark permits us to define Frobenius map on wave functions of B-model: using  (\ref {Frr1})-(\ref {Frr4}) and  the formula (8) from [12]  we see that we should
define  Frobenius map in canonical coordinates in the following way:
if ${\rm Fr}\psi =\psi '$ then
\begin{equation}
\label{s}
f'=f+  \big (\frac{1}{2\lambda ^2}(a\rho ^2+\delta a \rho+\frac{1}{2}\delta ^2 ax^2))
\end{equation}
As follows from (\ref {cf}) the Frobenius map  
preserves  the symplectic form on cohomology 
up to a constant factor, hence strictly speaking it is not 
symplectic map; nevertheless we can use the formulas from [12] to define its  action on wave functions. 
Notice that in (\ref {s})  $f'$ is calculated at the point $q$ and
$f$ at the point $q^p.$ 

We conjecture that the $p$-adic analog of the expression $\Psi$
obeys 
\begin{equation}
\label{P}
\frac{{\rm Fr}\Psi}{\Psi ^p}=1+p\zeta
\end{equation}
where $\zeta$ is an integral expression ( we understand integrality in the sense explained  after the formula (\ref {i})).
It follows from the above results that the integrality of Gopakumar-Vafa invariants is equivalent to  this
conjecture.

To verify this statement we  represent $\Psi$ as a product of three factors:
\begin{equation}
\label{ }
\Psi=\exp\big(F(\frac{\lambda}{\rho}, t+\frac{x}{\rho})\big)\exp \big(-{\lambda^{-2}\big(F_0(t)\rho^2+\frac{\partial
F_0}{\partial t}x\rho+\frac
{1}{2}\frac{\partial^2
F_0}{\partial t^2}x^2)\big)}\nonumber\\
\exp \big (-(\frac{\chi}{24}-1)\log(\rho)\big).
\end{equation}
It follows from our assumptions about Frobenius map and from
(\ref {f})
that  in
${\rm Fr}\Psi/ \Psi ^p$ the contribution of the second factor
in $\Psi$ cancels with the exponent coming from (\ref {s}).
We obtain that  (\ref {P}) is equivalent to (\ref  {i}).

It  follows from [1]  that in genus zero approximation our conjecture is true; it is easy to check this fact directly.

\section{Calculation of zeta-function}

In Sec 2 we calculated the Frobenius map on middle-dimensional cohomology in the neighborhood
of maximally unipotent boundary point. It was represented by means of a series with integral $p$-adic coefficients; this series
converges in the open unit $p$-adic disk. (For simplicity we assume again that $r=1.$)  However, changing a basis in cohomology in appropriate way we can enhance the domain where our formulas make sense. 

The right  basis in cohomology  consists of vectors 
$\Omega _0= \Omega, \Omega _1=\nabla _{\delta}\Omega ,\Omega _2= \nabla _{\delta}^2 \Omega, \Omega _3=\nabla _{\delta}^3 \Omega,$ where $\Omega$ stands for the cohomology class of the holomorphic 3-form. ( We assume that
$\Omega$ is a holomorphic function on the whole moduli space of complex structures.) We can write $\Omega = f e_0$ where $f(z)$ is a function 
holomorphic at $z=0.$ One can characterize $f(z)$ as a holomorphic solution of Picard-Fuchs equation. 
Applying  (\ref {gm1}-\ref {gm4}) we obtain
\begin{equation}
\Omega _0= fe_0, 
\end{equation}
\begin{equation}
\Omega _1= \delta f e_0+fY_1e_1,
\end{equation}
\begin{equation}
\Omega _2=\delta ^2f e_0+2\delta f Y_1e_1+f\delta Y_1e_1+fY_1Y_2e_2,
\end{equation}
\begin{equation}
\Omega  _3=\delta ^3f e_0+3\delta ^2fY_1e_1+3\delta f\delta Y_1e_1+f\delta ^2Y_1e_1+3\delta f Y_1 Y_2e_2+2f\delta Y_1Y_2e_2+fY_1\delta Y_2e_2+fY_1Y_2Y_3e_3.
\end{equation}

It is well known that zeta-function
can be  expressed in terms of traces on cohomology of  Frobenius map and its powers
at Teichmueller points. (Teichmueller  points are defined as 
elements of the field $\mathbb{C}_p$ of complex $p$-adic numbers obeying $z^{p^s}=z$
for some $s\in \mathbb{N}.$ In other words, these points are fixed points of powers of Frobenius map $z\to z^p.$ Only at Teichmueller point  the transformation ${\rm Fr}^{(s)}$ corresponding to a power of map $z\to z^p$  maps a cohomology group into itself and we can talk about the trace of this transformation.) 

To calculate the zeta-function we should know the Frobenius map on all cohomology groups. Our results are not sufficient for this calculation if we impose only the condition of existence of non-vanishing 
holomorphic  3-form in the definition of Calabi-Yau threefold. (Only this condition was necessary to derive the results of Sec 2 and 3.)  However, very often one
includes additional conditions in the definition of Calabi-Yau manifold $X$; namely
one requires that  holomorphic forms of positive degree  exist only in maximal
dimension (i.e. the Hodge numbers $h^{k0}$ vanish unless $k=0$ or $k=\dim X).$
If these conditions are included all  cohomology classes on Calabi-Yau threefold, except classes in middle-dimensional cohomology, can be represented by means of algebraic cycles.
If  an
algebraic cycle   is a manifold specified by equations with rational coefficients then at a Teichmueller point an element of cohomology group  dual to the cycle  is an eigenvector of  (a power of )
Frobenius map; the eigenvalue  of $F^s$ is
equal to $p^{ds}$ where $d$ stands for the codimension of algebraic cycle.
The proof of this fact is based on the remark that
one can define  Frobenius map of algebraic cycles that is Poincare dual to the Frobenius map on cohomology and that the Frobenius map on algebraic cycles of the submanifold agrees 
with the Frobenius map on algebraic cycles  of the whole manifold.
If the cohomology group is spanned by algebraic cycles given by rational equations
the action of  $Fr^s$ reduces to multiplication by 
$p^{ds}$.
This happens, in particular, for toric manifolds
and for complete intersections in toric manifolds
in the case when the restriction map of two-dimensional cohomology of toric manifold into
cohomology of complete intersection is surjective (for example, for quintic) . 

A cohomology class that corresponds to an  arbitrary algebraic subvariety  still is an eigenvector with respect to $Fr^s$, but the eigenvalue can be equal a root of unity multiplied
by $p^{ds}$.

This implies that the eigenvalues of the Frobenius  on the second cohomology group have the form $\epsilon_i p$, where $\epsilon_i$ is a root of unity, and
on the fourth cohomology group the eigenvalues are equal to $\epsilon_i^{-1} p^2$.

It is well known that zeta-function can be represented as as a quotient of 
two polynomials; odd-dimensional cohomology groups contribute to the
numerator $P(t)$ and even-dimensional cohomology groups contribute to the denominator $Q(t).$ In the situation at hand it is easy to calculate the 
denominator. 

If the cohomology group is spanned by  algebraic cycles given by rational equations one obtains
\begin{equation}
\label{q}
Q(t)=(1-t)(1-pt)^{h^{11}}(1-p^2t)^{h^{11}}(1-p^3t).
\end{equation}

In general case
\begin{equation}
\label{Q}
Q(t)=(1-t) (1-p^3t)  \,  \prod_{1\leq i \leq  h^{11}} (1-\epsilon_i pt)(1- \epsilon_i^{-1}p^2t) .
\end{equation} 
All odd-dimensional cohomology groups except $H^3$ vanish, therefore we can
obtain an expression of zeta-function in terms of Frobenius transformation on
middle-dimensional cohomology calculated in Sec 2.  We will not reproduce the
result of this trivial, but lengthy calculation. Similar calculations can be performed
in the case of elliptic curves (in this case we rederive  Dwork's result [9]) and in the case of K3 surfaces (Calabi-Yau manifolds of dimension 2). 

 To give a flavor of formulas obtained by means of the methods described above we consider  much simpler
problem: how to calculate the number of points  mod $p^2.$ (We are talking about the number of points of  Calabi-Yau manifold over
$\mathbb{F}_p$ . )  As in general case  we should calculate the trace of the operator ${\rm Fr}$ on the cohomology group over Teichmueller point using the basis  $\Omega _0= \Omega, \Omega _1=\nabla _{\delta}\Omega  ,\Omega _2= \nabla _{\delta}^2 \Omega, \Omega _3=\nabla _{\delta}^3 \Omega,$ but we can drop in our calculations all terms containing $p^2 .$  The calculation  of this trace leads to the following result:
\begin{equation}
\label{ }
{\rm Tr} {\rm Fr}=\hat{A}+\hat B p (1+\frac{a}{Y_1}({\rm Fr}(U)-U))
\end{equation}
where $a= a_{32}$ is a matrix element of ${\rm Fr}$ (see (\ref {Fr2})), $A=fY_1^2Y_2, B=fY_1Y_2$ ,$U= \delta \log (f^3Y_1^2Y_2).$ We use here the notation $$\hat R =\frac{{\rm Fr} (R)}{R}.$$
It seems that $U$ is well defined at Teichmueller point; if this is true then
${\rm Fr}(U)=U$ and
$${\rm Tr} {\rm Fr}=\hat{A}+\hat B p.$$
Notice that one can simplify this formula using (\ref {fyy}) . In particular, we obtain that  mod $p$
\begin{equation}
\label{A}
{\rm Tr} {\rm Fr}=\hat {A}=\widehat{f^{-1}}= \frac{f}{{\rm Fr} (f)}.
\end{equation}

The above considerations permit us to say that
in principle zeta-function of Calabi-Yau threefold
can be expressed in terms of mirror map, instanton numbers and holomorphic period $f$.

There exists also an expression of zeta function
in terms of all periods (solutions of Picard-Fuchs
equation). Recall, that the periods $f=f_0, f_1, f_2, f_3$ can be interpreted as integrals of the
form $\Omega$ over covariantly constant families of cycles or as scalar products with
covariantly constant forms $g_0,g_1,g_2,g_3$.
The  scalar products $<g_a, \Omega _b>=<g_a, \nabla ^b \Omega>$
can be obtained from periods by means of differentiation. We will express zeta-function in
terms of matrix $f_{ab}$ of these scalar products
(period matrix). Here $0\leq a,b\leq 3$. Let us 
assume that the forms $g_0,g_1,g_2,g_3$ constitute a symplectic basis in three-dimensional cohomology. Then   $\Omega _a= r_a^b g_b$ where the matrix $r$ is a product of
symplectic matrix $\varepsilon$ and period 
matrix $f$. The matrix, inverse to $r$, will be 
denoted by $\rho$.  
Periods are series with respect to $z, \log z$
having rational coefficients. Hence we can 
consider $p$-adic reduction of periods as well as
of matrices $r$ and $\rho$. (Recall that logarithm
makes sense in $p$-adic setting.) Therefore
starting with the basis $\Omega _i$ that makes
sense over $p$-adic numbers we can construct
a basis $g_i$ of $p$-adic cohomology, related to
$\Omega _i$ by the same formulas as in 
complex case. This basis also will be covariantly
constant.  It follows immediately from (\ref {c})
that the matrix of Frobenius map in covariantly
constant basis is constant; hence it is sufficient
to calculate it at the point $z=0$. The behavior of
Frobenius map at this point was discussed in
Sec. 2 on the base of results of [2]. 
 
We can say that the matrix of a power of Frobenius map ${\rm Fr}^{(s)}$ in
the basis $\Omega _i$ is a product of the matrix ${\rm Fr}^sr$, the constant matrix of ${\rm Fr}^{(s)}$  at $z=0$
and the matrix $\rho$. Calculating the trace of this matrix at Teichmueller points  we can obtain 
zeta-function. Notice, that the entries of 
period matrix are series  diverging
at Teichmueller points,but the entries of the
matrix of the power of Frobenius map are given
by series that converge at these points.

The zeta-function of quintic was analyzed in [8]. The formula (\ref {A}) agrees with [8].
It would be interesting to rederive other results of [8]  using our calculation; it seems that this is possible, but not simple. 

{\bf  Acknowledgments} 

We are indebted to M. Kontsevich , A. Ogus and Daqing Wan for interesting  discussions. 

Both authors were partially supported by NSF  grants 
(the first  author by DMS-0505735 and the second author by
DMS-0401164).

{\bf  References}

1. M. Kontsevich, A. Schwarz, V. Vologodsky, Integrality of
instanton numbers and $p$-adic B-model, hep-th/0603106,
Phys.Lett. B637 (2006) 97-101

2.  V. Vologodsky, On integrality
of instanton numbers, arXiv:0707.4617

3.A. Schwarz, I. Shapiro,  Supergeometry and Arithmetic Geometry,
hep-th/0605119,  $p$-adic superspaces and Frobenius, math.NT/0605310 

4.D. Cox, S. Katz, Mirror symmetry and algebraic geometry,
AMS, 1999

5. D. Morisson, Mirror symmetry and rational curves on quintic
threefolds: A guide for mathematicians, alg-geom/9202004;
Compactifications of moduli spaces inspired by mirror symmetry,alg-geom/9304007; Mathematical aspects of mirror symmetry, alg-geom/960921

6. B. Lian, S. T. Yau, Mirror maps, modular relations and hypergeometric series, hep-th/9507151

7. P. Deligne, Local behavior of Hodge structures at infinity,
in:Mirror Symmetry II (B. Greene and S.-T. Yau, eds)  AMS,
Providence,RI,1997, 683-699

8. P. Candelas, X. de la Ossa and F. Rodriguez-Villegas,
Calabi-Yau manifolds over finite fields, I, hep-th/0012233,II,hep-th/0402133 

9. B. Dwork,  Normalized Period Matrices I: Plane Curves,
Annals of Mathematics, 2nd Ser., Vol. 94, No. 2 (Sep., 1971) , pp. 337-388

10.Bershadsky, M., Cecotti, S., Ooguri, H., and Vafa, C.,
Kodaira-Spencer Theory of Gravity and Exact Results for Quantum
String Amplitude,  Comm. Math. Phys., 165 (1994), 211-408.

11.E. Witten , Quantum Background Independence In String Theory, hep-th/9306112.

12. A. Schwarz, Xiang Tang, Holomorphic anomaly and quantization, hep-th/0611280

13. Katz, S., Gromov-Witten, Gopakumar-Vafa, and Donaldson-Thomas
invariants of Calabi-Yau threefolds,  math.AG/0408266.

%\subsection{}

\end{document}